**Autism spectrum disorder: a neuro-immunometabolic hypothesis of the developmental origins**

Martin G. Frasch[1,2], Byung-Jun Yoon[3], Dario-Lucas Helbing[4,5], Gal Snir[1], Marta C. Antonelli[6,7], Reinhard Bauer[4]

[1] Dept. of Obstetrics and Gynecology, University of Washington, Seattle, WA, USA.
[2] Center on Human Development and Disability, University of Washington, Seattle, WA, USA
[3] Electrical and Computer Engineering, Texas A&M University, College Station, TX, USA
[4] Institute for Molecular Cell Biology, Jena University Hospital, Friedrich Schiller University, Jena, Germany
[5] Leibniz Institute on Aging, Fritz Lipmann Institute, 07745 Jena, Germany
[6] Instituto de Biología Celular y Neurociencia "Prof. Eduardo De Robertis", Facultad de Medicina, Universidad de Buenos Aires, Argentina.
[7] Technical University of Munich, Germany; Institute for Advanced Study (Lichtenbergstrasse 2 a, D-85748 Garching, Germany).

Corresponding author:
Martin G. Frasch
Department of Obstetrics and Gynecology
University of Washington
1959 NE Pacific St
Box 356460
Seattle, WA 98195
Phone: +1-206-543-5892
Fax: +1-206-543-3915
Email: mfrasch@uw.edu

## Abstract

Fetal neuroinflammation and prenatal stress (PS) may contribute to lifelong neurological disabilities. Astrocytes and microglia, among the brain's non-neuronal "glia" cell populations, play a pivotal role in neurodevelopment, predisposition to and initiation of disease throughout lifespan. One of the most common neurodevelopmental disorders manifesting between 1-4 years of age is the autism spectrum disorder (ASD). A pathological glial-neuronal interplay is thought to increase the risk for clinical manifestation of ASD in at-risk children, but the mechanisms remain poorly understood and integrative, multi-scale models are needed. We propose a model that integrates the data across the scales of physiological organization, from genome to phenotype, and provides a foundation to explain the disparate findings on the genomic level. We hypothesize that via gene-environment interactions, fetal neuroinflammation and PS may reprogram glial immunometabolic phenotypes that impact neurodevelopment and neurobehavior. Drawing on genomic data from the recently published series of ovine and rodent glial transcriptome analyses with fetuses exposed to neuroinflammation or PS, we conduct an analysis on the Simons Foundation Autism Research Initiative (SFARI) Gene database. We confirm 21 gene hits. Using unsupervised statistical network analysis, we then identify six clusters of probable protein-protein interactions mapping onto the immunometabolic and stress response networks and epigenetic memory. These findings support our hypothesis. We discuss the implications for ASD etiology, early detection, and novel therapeutic approaches. We conclude with delineation of the next steps to verify our model on the individual gene level in an assumption-free manner. The proposed model is of interest for the multidisciplinary community of stakeholders engaged in ASD research, the development of novel pharmacological and non-pharmacological treatments, early prevention, detection as well as for the policy makers.

## 1. Introduction

A body of preclinical and epidemiological evidence shows that *in utero* exposure to infection or stress results in altered neurodevelopmental trajectories including psychiatric conditions such as autism spectrum disorder (ASD).[1]

However, the mechanisms explaining the individual risk and framework for early detection and treatment have remained elusive. Here we synthesize the evidence of connections between fetal neuroinflammation, prenatal stress (PS) and ASD to generate a new model of ASD etiology accounting for the contribution of microglia and astrocytes, the key players of brain inflammation. We deploy a multi-species approach to data synthesis.[2] Combining biomarkers derived from observations in multiple species such as rodent, ovine and human in the present case, should increase the likelihood of identifying a translationally valid disease etiology.

As a result, we propose a novel model of neuro-immunometabolic etiology of ASD. By way of an initial validation of this model on the population level, in a large genome database sample of children affected by ASD, we test for the presence of disruptive mutations to genes involved in the respective pathways and systems. We conclude with discussion of the implications and recommendations for assumption-free validation of the proposed model in the existing genomic data on the individual level, the subject of an ongoing research. The proposed model is of interest for the multidisciplinary community of stakeholders engaged in ASD research, the development of novel pharmacological and non-pharmacological treatments, early prevention, detection as well as for the policy makers.

## 2. Methods

**Gathering the biomarkers across cell systems and species**

Based on our observations in fetal microglia [3,4], the simplistic hypothesis is that exposure to endotoxin or chronic stress augments energy conservation - driving pathways and the α7 nicotinic acetylcholine receptor α7nAChR agonism reduces that response. Such gene-environment adaptations should be reflected in changes of epigenetic memory mediated by histone acetylation/deacetylation enzymes (HDAC/HAT) involved in microglial memory of inflammation.[3]

Synthesizing findings from the studies in the rodent, ovine and human species, we identified genes associated with the inflammatory, stress and metabolic phenotype of microglia and astrocytes (Fig. 1 and Table 1). We included the genes involved in energy homeostasis based on the observations linking neuroimmune and metabolic memory signatures in fetal microglia exposed to LPS and the larger emerging gestalt of physiological stress response in immunometabolism, in particular by astrocytes.[5,6]

We included the iron homeostasis genes based on the recently postulated interaction between this signaling network and the α7nAChR modulation of immunometabolic phenotype of fetal microglia.[4]

We included the complement pathway genes expressed in microglia and known to be modulated by neuroinflammation in microglia and to play an important role in neurodevelopmental synaptic pruning.[4,7]

Based on the evidence that *in utero* inflammation or stress increase risk for ASD, we sought to identify genes linked to the ASD diagnosis and falling into the family of signaling pathways listed in Table 1. We tested that hypothesis statistically against the Simons Foundation Autism Research Initiative (SFARI) Gene database. The SFARI Gene is an evolving online database for the autism research community that is centered on genes implicated in autism susceptibility.

Figure 1 summarizes the systems level approach to selecting the targets from animal models and identifying those that are also listed in SFARI. Such targets are listed centered in each system's box. We provide further details on the selected targets in Table 1.

**Table 1. Genes implicated in the signature of astrocytes and microglia exposed to endotoxin or stress**

| Groups | Genes | Function |
|---|---|---|
| Glial cell phenotype | *TMEM119* | Transmembrane protein 119; identifies resident microglia (from blood-derived macrophages) [8,9] |
| | ***TGFβ*** **[4]** | Transforming growth factor beta-1; resident microglial biomarker [10,11] |
| | *CD11b* | Resident microglial biomarker [9–11] |
| | *CD11c* | Activated microglia [10,11] |
| | ***CD45*** **[4]** | Protein Tyrosine Phosphatase Receptor Type C (PTPRC); Resident microglial biomarker [9–11] |
| | *Iba1* | Ionized calcium binding adaptor molecule 1: Non-specific microglial / macrophage biomarker [3,9–12] |
| | ***CX3CR1*** **[4]** | CX3C chemokine receptor 1; required for synaptic pruning during brain development [13] |
| | ***BRD4*** **[4]** | Bromodomain containing 4; polarizes microglia toward inflammatory phenotype [14]; involved in epigenetic memory [15] |
| | ***SLC1A2*** **[S]** | Astrocytic GLT-1 transporter required for neuron-astrocyte communication and astrocyte maturation [16] |
| Inflammation | *HMGB1* | Hypermobility group box protein 1, a pleiotropic signaling molecule in glia cells and neurons: a growth factor, a pro-inflammatory molecule [17]; implicated in ASD [18] |
| | *IL-10* | Key regulator of neuroimmune homeostasis via cross-talk of microglia and astrocytes [19] |
| | ***IL-6*** **[5]** | Early inflammatory cytokine |
| | *NLRP3* | Inflammasome activated in ASD [20] |
| Stress | ***CRH*** **[5]** | Corticotropin-releasing hormone, key hormone linking chronic stress with anxiety [21,22]; has direct effects on microglia [23] |
| | ***CRHR2*** **[5]** | Corticotropin-releasing hormone receptor 2; [23] |
| | ***HSD11B1*** **[4]** | 11β-Hydroxysteroid dehydrogenase type 1 [24] |
| | *POMC* | Proopiomelanocortin [25] |
| | *p-Akt* | Phosphorylated-Akt [26] |

| | ***PI3K*** **[S]** | PI3K/Akt signaling pathway; e.g. PIK3CA [26,27] |
|---|---|---|
| | ***OGT*** **[5]** | O-GlcNAc transferase; a placental biomarker of maternal stress exposure related to neurodevelopmental outcomes [28,29] |
| | ***GAP43*** **[5]** | Growth associated protein 43; in astrocytes, GAP43 mediates glial and neuronal plasticity during astrogliosis and attenuates microglial activation under LPS exposure [30] |
| | ***SLC22A3*** **[5]** | Solute carrier family 22 member 3; modulates anxiety and social interaction [31,32] |
| | ***PLPPR4*** **[5]** | Phospholipid Phosphatase Related 4; stress-related behaviors such as reduced resilience [33] |
| | ***PRKCB*** **[3]** | Protein Kinase C Beta; involved in stress-related behavior [34] |
| | ***UCN3*** **[5]** | Urocortin 3; binds specifically CRHR2 [23] |
| | ***DLG4*** **[5]** | Disks large homolog 4; modulates stress reactivity and anxiety [35,36] |
| Energy homeostasis | *Cx43* | Connexin 43 gap junction maintaining astrocytes' homeostasis via metabolic cooperation [37] |
| | *AMPK* | Adenosine monophosphate kinase, intracellular energy sensor[38] |
| | *FBP* | Fructo-biphosphokinase: signature of second-hit memory of inflammation in fetal microglia [3] |
| | ***mTOR*** **[S]** | Mammalian target of the rapamycin signaling pathway [39,40] |
| Iron homeostasis | *HAMP* | Hepcidin, a regulator of iron homeostasis and inflammation; implicated, along with ferroportin and transferrin, in microglial response to endotoxin interfering with α7nAChR signaling [4] |
| | *SLC40A1* | Ferroportin |
| | *TFR2* | Transferrin receptor 2; involved in iron sequestration |
| | *TFRC* | Transferrin receptor protein 1; needed for iron sequestration |
| | *HMOX1* | Hemoxygenase 1, key enzyme of iron homeostasis also serves as a signature of second-hit memory in fetal microglia and promoted by α7nAChR stimulation [3,41] |
| | ***SLC25A39*** **[4]** | Member of the SLC25 transporter or mitochondrial carrier family of proteins; required for normal heme biosynthesis |
| Complement pathway | *C1QA* | Aside from its traditional role in innate immunity [42], elements of the complement pathway are involved in neuronal-glial interactions; recognized as essential players in brain development, especially in synaptogenesis/synaptic pruning and predisposition for neurodegenerative diseases [7,43], their microglial expression is also susceptible to LPS exposure *in utero [4]* |
| | *C1QB* | |
| | *C[4]3AR1* | |
| | *CR2* | |
| | ***C4B*** **[4]** | |
| Epigenetic memory | ***HDAC*** **[S]** and ***HAT*** families | Histone acetylation/deacetylation enzymes: HDAC 1, 2, 3, 7 and 9 found in SSC database of ASD gene mutations. |

| | | HDAC 1, 2, **4** and 6 involved in fetal microglial memory of LPS exposure; HDAC 1, 2, **4** and 7 increased with altered neuronal AChE signaling and increased anxiety behavior due to adult chronic stress exposure [3,4,25,27,44,45] |
|---|---|---|

Hits in the SFARI database are in bold. ASD association score indicated in square brackets: S: syndromic; 1: high confidence; 2: strong candidate; 3: suggestive evidence; 4: minimal evidence; 5: hypothesized; 6: not supported. For details see https://gene-archive.sfari.org/about-gene-scoring/

"Syndromic" distinction contains mutations that are associated with a substantial degree of increased risk and consistently linked to additional characteristics not required for an ASD diagnosis.

We used that database to identify genes with varying degrees of disruptive mutations associated with ASD as defined in the SFARI database to any of the genes in Table 1. The analysis is based on accessing the database in August 2019 corresponding to the SFARI database update of June 20, 2019.

We identified 21 hits indicated in Table 1 in bold along with the level of confidence for association with ASD using the SFARI scoring criteria. Scores denoted as S identified syndromic category and lower numbers in the range from 1 to 6 mean higher confidence in the association, 1 being the highest and 6 being the lowest. It is apparent that all categories defined in Table 1 are represented in the SFARI database: microglial or astrocytic phenotype, inflammation, stress, energy homeostasis, iron homeostasis, complement pathway, and epigenetic memory.

## 3. Results

### Systems validation: many species, one network

We sought to validate the significance of the association of the identified protein products using network analysis with string-db.org. The results can be accessed online at the permanent URL. They represent the analysis of the subset of 21 proteins identified in Table 1. While individually 16 out of the 21 identified proteins were scored as weakly associated with ASD (scores 4 or 5 denoting minimal evidence or hypothesized), the network analysis shows a significant interaction between all 21 nodes on the protein-protein interaction level with six network clusters shown in Figure 1. These clusters form stress, metabolism, glia phenotype and epigenetic memory aspects of the proposed neuro-immunometabolic network exposed to stress or inflammation.

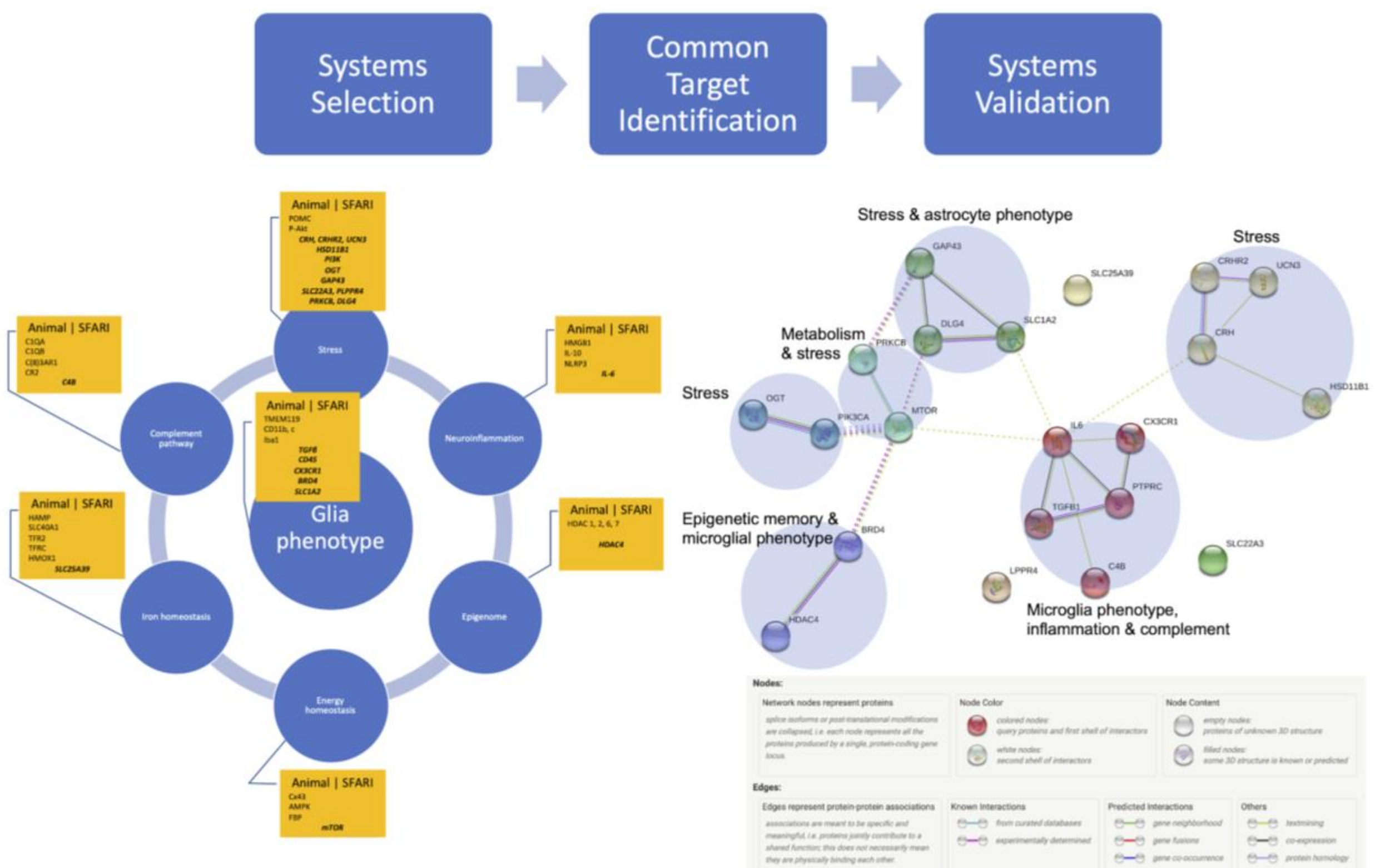

**Figure 1*.** Systems driven selection of potential targets from animal studies, followed by confirmation via SFARI database (centered in each box) thus identifying possible systems targets reflecting ASD (see Table 1 for details). Systems validation follows using the protein network analysis of ASD associated genes driven by exposure to stress or inflammation followed by Markov Cluster Algorithm for unsupervised clustering identifies significant interactions between six clusters containing microglial and astrocytic phenotypes, inflammation, stress, energy (but not iron) homeostasis, and complement pathway. Note that stress pathways are represented in four clusters (yellow, green, aquamarine, dark cyan), while microglial phenotype, inflammation and complement pathway comprise one cluster (red). Protein-protein interaction (PPI) enrichment p-value: 0.00085. PPI legend by string-db.org [46] The detailed list of participating node members is provided in the Supplementary Table S1.[47]

The data summarized in Figure 1 support our initial hypothesis linking both neuroinflammation and stress exposures *in utero* to the etiology of ASD in a gene-environment paradigm: developmental combinations of genetic alterations in key pathways susceptible to environmental stimuli of inflammation or stress increase risk for or cause ASD. The clusters of interactions map onto the neuro-immunometabolic and stress response networks.

## 4. Synthesis

### Information processing, energy demand and stress

Current estimates suggest that variation in as many as 1,000 different genes could affect susceptibility to ASD. In an attempt to tie this wealth of data together conceptually in a unified model, an endophenotype of autism has been proposed.[48] It is defined as a predictive impairment.

Sinha *et al.* note in [48] that providing a cohesive conceptualization of ASD, such as the predictive impairment endophenotype, would ameliorate the search for broadly effective therapies, diagnostic markers, and neural/genetic correlates. Indeed, setting the ASD on the information-theoretical platform in terms of prediction impairment reveals links to a major disruptor of neural information processing - the stress. For example, the prediction impairment hypothesis interprets the reduced habituation and hence greater stress in ASD subjects as being caused by an endogenous predictive impairment which leads the environmental stimuli to appear more chaotic. By the same token, an autistic individual responds to the chaotic-appearing environment with endogenous mechanisms geared to reducing the chaos by exhibiting ritualized behaviors. These features of the proposed ASD endophenotype all have in common an attempt to reduce stress arising from increased uncertainty.

Peters *et al.* provide a neurobiological perspective on these psychological observations.[49] Their observations are based on the well-founded physical-mathematical concepts linking the energy expenditure, entropy and information processing, going back to Shannon and Boltzmann.[50,51] This is combined with a body of neurobiological literature on the neuronal substrate of the brain's stress and prediction networks.[49] Herein, the so-called Selfish Bayesian Brain resolves stress, or information uncertainty, by an internal Bayesian update of the knowledge state. This must take place at the expense of transiently increased energy needs (hence "selfish"). If this update succeeds, the stress is a eustress; if it does not, it becomes a distress. The present findings suggest that one possible mechanism for when such an update can fail is when the increased energy demands cannot be met.

## Defining a cohesive cell systems correlate of ASD endophenotype

We are proposing a developmental origins perspective on this framework. The neuro-immunometabolic and stress response networks interact and *in utero* exposure to stress or inflammation may alter the glial immunometabolic phenotype, i.e., the pattern of this network interaction. That consequence translates into less energy availability under the increased demands due to stress.[49] If the energy needs are not met, the allostatic load turns into an allostatic overload.[52] Neuropathophysiological state may ensue unless the situation is corrected.

The above findings and their interpretation are further supported by the emerging evidence for the dynamic relationship between epigenetic memory and energy availability. In a broad sense, epigenetics refers to the chemical mechanisms that modify gene expression, without altering the DNA sequence. The main epigenetic modifications that change the chromatin architecture include DNA methylation, histone post-translational modifications and micro-RNA-mediated repression. While epigenetic regulation is dynamic, it is often also referred to as epigenetic memory, implying the stable propagation of changes in gene expression induced by a developmental or environmental stimulus.[53] Environmental stimuli include nutrients and energy metabolites to which we are exposed regularly and which are known to regulate gene expression through modulation of the epigenome. This paradigm was recently termed "metaboloepigenetics" bringing forward the crucial role of metabolites as co-factors of chromatin-modifying enzymes that are responsible for epigenetic modifications.[54,55] Enzymes such as methyltransferases, deacetylases and kinases employ cofactors such as α-ketoglutarate, acetyl-CoA, nicotinamide adenine dinucleotide, S-adenosylmethionine and ATP and the availability of these cofactors as well as nutrients such as glucose, oxygen and glutamine can reduce or enhance gene expression.[56] This implies that the response to a stressful microenvironment, such as intrauterine or postnatal adversity, will ultimately lead to a pathological state. Exposure to in utero stress and/or inflammation is likely to result in impaired levels of micronutrients and cofactors leading to a pathological methylation/demethylation of DNA and histone proteins and histone acetylation/deacetylation of the genes clustered in the neuro-immunometabolic network proposed. The presence of such effects can be validated in future preclinical and clinical studies.

## 5. Discussion

The proposed neuro-immunometabolic hypothesis can be used in future *in vivo* studies to probe causally for the role of the immunometabolic gene and protein networks in ASD etiology, in particular in the areas described as being the core of both the prediction processing network and contextual fear memory, the anterior cingulate cortex.[57–59] Indeed, a recent study reported that an immunometabolic dysregulation reduces the thickness of the anterior cingulate cortex, while another study revealed 1,223 differentially methylated genes (5,018 Differentiated Methylated Regions, DMRs) in this brain region 4 weeks after contextual fear conditioning.[60,61] In another most recent study in a human cohort of stressed pregnant mothers and their newborns, we identified novel associations between newborn epigenome-wide methylation status in saliva and chronic stress suffered by the mother during pregnancy.[62] Among the positively associated CpGs is the CpG annotated to ***YAP1*** (Yes1 regulated transcription factor) whose deletion has been related to reactive astrogliosis and astrocyte-driven microglial activation.[63] The protein of another gene (***CSMD1***, CUB And Sushi Multiple Domains 1) associated to maternal cortisol, CSMD1, seems to play a crucial role in regulating complement activation and inflammation in the developing brain.[64,65] Two DMRs were also found annotated to ***DAXX*** (Death-associated protein 6) and ***ARL4D*** (ADP- Ribosylation factor 4D), which together with ***CSMD1*** have all been directly or indirectly involved in neurodevelopmental disorders such as ASD, highlighting the impact of prenatal stress on the epigenetic landscape of the newborn in genes associated to neuroinflammation and ASD.[66–70]

Moreover, inflammatory priming in maternal immune activation (MIA) in general, can substantially alter neurodevelopment.[71] Specifically, microglia as the key players in MIA can modify neurogenesis, maturation of synapses and neural circuits. MIA due to various triggers, bacterial/viral or autoimmune triggers, as reviewed extensively elsewhere,[71] can influence the efficacy of microglia to sculpt neuronal connections. This can result in a dysfunctional neurodevelopment culminating in ASD and other neurodevelopmental disorders in the offspring such as attention-deficit/hyperactivity disorder and Tourette syndrome. Hence, inclusion of MIA studies might lead to additional insights into the role of neuro-immunometabolic changes in neurodevelopmental disorders such as ASD.

The therapeutic approaches may include prenatal and early neonatal screening using early biomarkers of altered brain development due to intrauterine exposure to stress or inflammation, followed by early postnatal reprogramming of these effects using behavioral (e.g., environmental enrichment) or bioelectronic tools to correct brain developmental trajectories at an early stage.[2,72–74]

In the context of the discussed neuro-immunometabolic and Selfish Bayesian Brain paradigms, environmental enrichment may help reduce the allostatic load by providing opportunities to perform the required knowledge state update in the face of existing uncertainties, i.e., stress.

Other salutary mechanisms mediating the therapeutic approaches may involve activation of the afferent cholinergic pathway. Vagal afferent signaling affects widely distributed areas of the brain via Nucleus tractus solitarii, and studies indicate that stimulation of the vagus nerve activates many neurons in both cerebral hemispheres and hippocampus.[75,76] Vagal signaling in near-term fetus decreases HMGB1 concentrations in neuronal and astrocyte cytoplasm expressing α7nAChR and decreases microglial activation, which may decrease glial priming.[12] Hence, efferent peripheral or afferent vagal activity may mediate central neuroprotection via regulation of neuroinflammation. Vagus nerve stimulation has been shown to alter the phase synchrony in the anterior cingulate cortex improving decision making in rats.[77]

The efferent wiring of the vagal cholinergic pathway has begun to unravel in the last decade, while our understanding of the afferent signaling is only beginning to emerge. The efferent component is also referred to as the inflammatory reflex which acts via the cholinergic anti-inflammatory pathway (CAP), a neural mechanism that influences the magnitude of innate immune responses to inflammatory stimuli and maintains homeostasis.[78] Through CAP, increased vagal activity inhibits the release of pro-inflammatory cytokines. We found that spontaneous CAP activity is present in the sheep fetus as early as at 0.76 gestation.[79] Moreover, CAP activation near-term suppresses activation of ovine microglia and astrocytes which express α7 nAChR or hypoxic-ischemic brain damage in newborn rats.[12,17,80–83]

## 6. Conclusions

The proposed model suggests that early behavioral therapy or enhancing fetal CAP activity via vagal nerve stimulation or α7nAChR agonists will suppress the activation of microglia and astrocytes restoring their physiological immunometabolic phenotype, thus decreasing glial priming and preventing a sustained switch to a reactive phenotype. This approach might decrease the degree of brain injury sustained from fetal systemic and brain inflammatory responses and stress exposures, thus improving postnatal short- and long-term health outcomes, including susceptibility to ASD.

Our proposed model and hypothesis presented in this study may be further strengthened and deepened through validation via a model-free approach. This can be achieved by analyzing the individual genomic and transcriptomic information from the same SFARI database and evaluating the evidence of the presence of the observed networks - by investigating independent datasets that were not utilized in deriving the networks in Figure 1 by means of methodologically distinct analytic approaches. For this purpose, we envision the use of the following three computational pathway and network analysis schemes. First, we may quantitatively assess the relevance and relative importance of known pathways (e.g., canonical pathways curated in KEGG database) based on the differential activity of each pathway under different conditions (e.g., whether or not an individual has experienced exposure to fetal neuroinflammation and prenatal stress). It has been shown that differential pathway activity can be effectively inferred from gene expression data using probabilistic graphical models, and that highly differentially activated pathways detected therefrom can serve as better biomarkers for complex diseases (such as cancer) with higher discriminative power as well as improved reproducibility.[84] Next, we may identify differentially expressed genes or differentially activated pathways as “seeds” and identify potential network modules that include these seeds. For example, we may adapt the strategy proposed by Wang et al. [85], in order to find out whether the seed genes/pathways would be part of a well-conserved network modules with characteristic topological features often sighted in known functional modules (e.g., signaling pathways, protein complexes). Another promising approach is to carry out a network-based analysis of the transcriptomic data in integration with protein interaction data for unsupervised detection of potential network modules that may be associated with the cellular response to fetal neuroinflammation and prenatal stress. As shown in [86], recent advances in deep learning (esp., graph convolutional networks) can facilitate such investigation. These remain the subjects of our ongoing research.

## Data accessibility

The data is citable as [47] and accessible on GitHub:
https://github.com/martinfrasch/ASD_origins_hypothesis

## Authors' contributions

MGF conceived the manuscript and wrote the initial draft. GS contributed to the data analysis. MCA, AS, DLH and RB contributed to the manuscript. All authors gave final approval for publication and agree to be held accountable for the work performed therein.

## Competing interests

We declare that we have no competing interests.

## Funding

None.

## Acknowledgements

The authors thank Raphael A. Bernier and Elisabeth A. Rosenthal for helpful advice on the ASD and genetics aspects of the secondary analysis.

## References

1. al-Haddad BJS *et al.* 2019 Long-term Risk of Neuropsychiatric Disease After Exposure to Infection In Utero. *JAMA Psychiatry*. (doi:10.1001/jamapsychiatry.2019.0029)

2. Frasch MG *et al.* 2018 Non-invasive biomarkers of fetal brain development reflecting prenatal stress: An integrative multi-scale multi-species perspective on data collection and analysis. *Neurosci. Biobehav. Rev.* (doi:10.1016/j.neubiorev.2018.05.026)

3. Cao M *et al.* 2015 Fetal microglial phenotype in vitro carries memory of prior in vivo exposure to inflammation. *Front. Cell. Neurosci.* **9**, 294. (doi:10.3389/fncel.2015.00294)

4. Cortes M *et al.* 2017 α7 nicotinic acetylcholine receptor signaling modulates the inflammatory phenotype of fetal brain microglia: first evidence of interference by iron homeostasis. *Sci. Rep.* **7**, 10645. (doi:10.1038/s41598-017-09439-z)

5. Lee YS, Wollam J, Olefsky JM. 2018 An Integrated View of Immunometabolism. *Cell* **172**, 22–40. (doi:10.1016/j.cell.2017.12.025)

6. Camandola S. 2018 Astrocytes, emerging stars of energy homeostasis. *Cell Stress Chaperones* **2**, 246–252. (doi:10.15698/cst2018.10.157)

7. Schafer DP *et al.* 2012 Microglia sculpt postnatal neural circuits in an activity and complement-dependent manner. *Neuron* **74**, 691–705. (doi:10.1016/j.neuron.2012.03.026)

8. Satoh J-I, Kino Y, Asahina N, Takitani M, Miyoshi J, Ishida T, Saito Y. 2016 TMEM119 marks a subset of microglia in the human brain. *Neuropathology* **36**, 39–49. (doi:10.1111/neup.12235)

9. Bennett ML *et al.* 2016 New tools for studying microglia in the mouse and human CNS. *Proc. Natl. Acad. Sci. U. S. A.* **113**, E1738–46. (doi:10.1073/pnas.1525528113)

10. Butovsky O *et al.* 2014 Identification of a unique TGF-beta-dependent molecular and functional signature in microglia. *Nat. Neurosci.* **17**, 131–143. (doi:10.1038/nn.3599)

11. Abud EM *et al.* 2017 iPSC-Derived Human Microglia-like Cells to Study Neurological Diseases. *Neuron* **94**, 278–293.e9. (doi:10.1016/j.neuron.2017.03.042)

12. Frasch MG, Szynkaruk M, Prout AP, Nygard K, Cao M, Veldhuizen R, Hammond R, Richardson BS. 2016 Decreased neuroinflammation correlates to higher vagus nerve activity fluctuations in near-term ovine fetuses: a case for the afferent cholinergic anti-inflammatory pathway? *J. Neuroinflammation* **13**, 103. (doi:10.1186/s12974-016-0567-x)

13. Paolicelli RC *et al.* 2011 Synaptic pruning by microglia is necessary for normal brain development. *Science* **333**, 1456–1458. (doi:10.1126/science.1202529)

14. Wang J *et al.* 2019 BRD4 inhibition attenuates inflammatory response in microglia and

facilitates recovery after spinal cord injury in rats. *J. Cell. Mol. Med.* **23**, 3214–3223. (doi:10.1111/jcmm.14196)

15. Penas C, Navarro X. 2018 Epigenetic Modifications Associated to Neuroinflammation and Neuropathic Pain After Neural Trauma. *Front. Cell. Neurosci.* **12**, 158. (doi:10.3389/fncel.2018.00158)

16. Hasel P *et al.* 2017 Neurons and neuronal activity control gene expression in astrocytes to regulate their development and metabolism. *Nat. Commun.* **8**, 15132. (doi:10.1038/ncomms15132)

17. Frasch MG, Nygard KL. 2017 Location, Location, Location: Appraising the Pleiotropic Function of HMGB1 in Fetal Brain. *J. Neuropathol. Exp. Neurol.* **76**, 332–334. (doi:10.1093/jnen/nlx004)

18. Dipasquale V, Cutrupi MC, Colavita L, Manti S, Cuppari C, Salpietro C. 2017 Neuroinflammation in Autism Spectrum Disorders: Role of High Mobility Group Box 1 Protein. *Int J Mol Cell Med* **6**, 148–155. (doi:10.22088/acadpub.BUMS.6.3.148)

19. Lobo-Silva D, Carriche GM, Castro AG, Roque S, Saraiva M. 2016 Balancing the immune response in the brain: IL-10 and its regulation. *J. Neuroinflammation* **13**, 297. (doi:10.1186/s12974-016-0763-8)

20. Saresella M *et al.* 2016 Multiple inflammasome complexes are activated in autistic spectrum disorders. *Brain Behav. Immun.* **57**, 125–133. (doi:10.1016/j.bbi.2016.03.009)

21. Tsilioni I, Dodman N, Petra AI, Taliou A, Francis K, Moon-Fanelli A, Shuster L, Theoharides TC. 2014 Elevated serum neurotensin and CRH levels in children with autistic spectrum disorders and tail-chasing Bull Terriers with a phenotype similar to autism. *Transl. Psychiatry* **4**, e466. (doi:10.1038/tp.2014.106)

22. Theoharides TC, Asadi S, Patel AB. 2013 Focal brain inflammation and autism. *J. Neuroinflammation* **10**, 46. (doi:10.1186/1742-2094-10-46)

23. Kato TA, Hayakawa K, Monji A, Kanba S. 2013 Missing and Possible Link between Neuroendocrine Factors, Neuropsychiatric Disorders, and Microglia. *Front. Integr. Neurosci.* **7**, 53. (doi:10.3389/fnint.2013.00053)

24. Wyrwoll CS, Holmes MC, Seckl JR. 2011 11β-hydroxysteroid dehydrogenases and the brain: from zero to hero, a decade of progress. *Front. Neuroendocrinol.* **32**, 265–286. (doi:10.1016/j.yfrne.2010.12.001)

25. Lemche E, Chaban OS, Lemche AV. 2016 Neuroendorine and Epigentic Mechanisms Subserving Autonomic Imbalance and HPA Dysfunction in the Metabolic Syndrome. *Front. Neurosci.* **10**, 142. (doi:10.3389/fnins.2016.00142)

26. Yang P-C, Yang C-H, Huang C-C, Hsu K-S. 2008 Phosphatidylinositol 3-kinase activation

is required for stress protocol-induced modification of hippocampal synaptic plasticity. *J. Biol. Chem.* **283**, 2631–2643. (doi:10.1074/jbc.M706954200)

27. Shi H-S *et al.* 2012 PI3K/Akt signaling pathway in the basolateral amygdala mediates the rapid antidepressant-like effects of trefoil factor 3. *Neuropsychopharmacology* **37**, 2671–2683. (doi:10.1038/npp.2012.131)

28. Howerton CL, Morgan CP, Fischer DB, Bale TL. 2013 O-GlcNAc transferase (OGT) as a placental biomarker of maternal stress and reprogramming of CNS gene transcription in development. *Proc. Natl. Acad. Sci. U. S. A.* **110**, 5169–5174. (doi:10.1073/pnas.1300065110)

29. Howerton CL, Bale TL. 2014 Targeted placental deletion of OGT recapitulates the prenatal stress phenotype including hypothalamic mitochondrial dysfunction. *Proc. Natl. Acad. Sci. U. S. A.* **111**, 9639–9644. (doi:10.1073/pnas.1401203111)

30. Hung C-C *et al.* 2016 Astrocytic GAP43 Induced by the TLR4/NF-κB/STAT3 Axis Attenuates Astrogliosis-Mediated Microglial Activation and Neurotoxicity. *J. Neurosci.* **36**, 2027–2043. (doi:10.1523/JNEUROSCI.3457-15.2016)

31. Wultsch T *et al.* 2009 Decreased anxiety in mice lacking the organic cation transporter 3. *J. Neural Transm.* **116**, 689–697. (doi:10.1007/s00702-009-0205-1)

32. Garbarino VR, Santos TA, Nelson AR, Zhang WQ, Smolik CM, Javors MA, Daws LC, Gould GG. 2019 Prenatal metformin exposure or organic cation transporter 3 knock-out curbs social interaction preference in male mice. *Pharmacol. Res.* **140**, 21–32. (doi:10.1016/j.phrs.2018.11.013)

33. Vogt J *et al.* 2016 Molecular cause and functional impact of altered synaptic lipid signaling due to a prg-1 gene SNP. *EMBO Mol. Med.* **8**, 25–38. (doi:10.15252/emmm.201505677)

34. Lisowski P, Wieczorek M, Goscik J, Juszczak GR, Stankiewicz AM, Zwierzchowski L, Swiergiel AH. 2013 Effects of chronic stress on prefrontal cortex transcriptome in mice displaying different genetic backgrounds. *J. Mol. Neurosci.* **50**, 33–57. (doi:10.1007/s12031-012-9850-1)

35. Feyder M *et al.* 2010 Association of mouse Dlg4 (PSD-95) gene deletion and human DLG4 gene variation with phenotypes relevant to autism spectrum disorders and Williams' syndrome. *Am. J. Psychiatry* **167**, 1508–1517. (doi:10.1176/appi.ajp.2010.10040484)

36. Li J, Shi M, Ma Z, Zhao S, Euskirchen G, Ziskin J, Urban A, Hallmayer J, Snyder M. 2014 Integrated systems analysis reveals a molecular network underlying autism spectrum disorders. *Mol. Syst. Biol.* **10**, 774. (doi:10.15252/msb.20145487)

37. Contreras JE, Sánchez HA, Eugenin EA, Speidel D, Theis M, Willecke K, Bukauskas FF, Bennett MVL, Sáez JC. 2002 Metabolic inhibition induces opening of unapposed connexin 43 gap junction hemichannels and reduces gap junctional communication in cortical astrocytes in culture. *Proc. Natl. Acad. Sci. U. S. A.* **99**, 495–500. (doi:10.1073/pnas.012589799)

38. Frasch MG. 2014 Putative Role of AMPK in Fetal Adaptive Brain Shut-Down: Linking Metabolism and Inflammation in the Brain. *Front. Neurol.* **5**, 150. (doi:10.3389/fneur.2014.00150)

39. Hui K, Katayama Y, Nakayama KI, Nomura J, Sakurai T. 2018 Characterizing vulnerable brain areas and circuits in mouse models of autism: Towards understanding pathogenesis and new therapeutic approaches. *Neurosci. Biobehav. Rev.* (doi:10.1016/j.neubiorev.2018.08.001)

40. Wang H, Doering LC. 2013 Reversing autism by targeting downstream mTOR signaling. *Front. Cell. Neurosci.* **7**, 28. (doi:10.3389/fncel.2013.00028)

41. Hua S, Ek CJ, Mallard C, Johansson ME. 2014 Perinatal hypoxia-ischemia reduces alpha 7 nicotinic receptor expression and selective alpha 7 nicotinic receptor stimulation suppresses inflammation and promotes microglial Mox phenotype. *Biomed Res. Int.* **2014**, 718769. (doi:10.1155/2014/718769)

42. Noris M, Remuzzi G. 2013 Overview of complement activation and regulation. *Semin. Nephrol.* **33**, 479–492. (doi:10.1016/j.semnephrol.2013.08.001)

43. Hong S *et al.* 2016 Complement and microglia mediate early synapse loss in Alzheimer mouse models. *Science* **352**, 712–716. (doi:10.1126/science.aad8373)

44. Thorsell A. 2010 Brain neuropeptide Y and corticotropin-releasing hormone in mediating stress and anxiety. *Exp. Biol. Med.* **235**, 1163–1167. (doi:10.1258/ebm.2010.009331)

45. Sailaja BS, Cohen-Carmon D, Zimmerman G, Soreq H, Meshorer E. 2012 Stress-induced epigenetic transcriptional memory of acetylcholinesterase by HDAC4. *Proc. Natl. Acad. Sci. U. S. A.* **109**, E3687–95. (doi:10.1073/pnas.1209990110)

46. Szklarczyk D *et al.* 2019 STRING v11: protein-protein association networks with increased coverage, supporting functional discovery in genome-wide experimental datasets. *Nucleic Acids Res.* **47**, D607–D613. (doi:10.1093/nar/gky1131)

47. Frasch M. 2019 martinfrasch/ASD_origins_hypothesis 1. (doi:10.5281/zenodo.3402349)

48. Sinha P, Kjelgaard MM, Gandhi TK, Tsourides K, Cardinaux AL, Pantazis D, Diamond SP, Held RM. 2014 Autism as a disorder of prediction. *Proc. Natl. Acad. Sci. U. S. A.* **111**, 15220–15225. (doi:10.1073/pnas.1416797111)

49. Peters A, McEwen BS, Friston K. 2017 Uncertainty and stress: Why it causes diseases and how it is mastered by the brain. *Prog. Neurobiol.* **156**, 164–188. (doi:10.1016/j.pneurobio.2017.05.004)

50. Shannon CE. 1948 A Mathematical Theory of Communication. *Bell System Technical Journal* **27**, 379–423. (doi:10.1002/j.1538-7305.1948.tb01338.x)

51. Boltzmann ML. 1877 Über die Beziehung zwischen dem zweiten Hauptsatze des mechanischen Wärmetheorie und der Wahrscheinlichkeitsrechnung, respective den Sätzen über das Wärmegleichgewicht. On the relationship between the second main theorem of mechanical heat theory and the probability calculation with respect to the results about the heat equilibrium. *Wiener Berichte* **76**, 373–435.

52. McEwen BS, Stellar E. 1993 Stress and the individual. Mechanisms leading to disease. *Arch. Intern. Med.* **153**, 2093–2101.

53. D'Urso A, Brickner JH. 2014 Mechanisms of epigenetic memory. *Trends Genet.* **30**, 230–236. (doi:10.1016/j.tig.2014.04.004)

54. Donohoe DR, Bultman SJ. 2012 Metaboloepigenetics: interrelationships between energy metabolism and epigenetic control of gene expression. *J. Cell. Physiol.* **227**, 3169–3177. (doi:10.1002/jcp.24054)

55. Etchegaray J-P, Mostoslavsky R. 2016 Interplay between Metabolism and Epigenetics: A Nuclear Adaptation to Environmental Changes. *Mol. Cell* **62**, 695–711. (doi:10.1016/j.molcel.2016.05.029)

56. Wang Z, Long H, Chang C, Zhao M, Lu Q. 2018 Crosstalk between metabolism and epigenetic modifications in autoimmune diseases: a comprehensive overview. *Cell. Mol. Life Sci.* **75**, 3353–3369. (doi:10.1007/s00018-018-2864-2)

57. Thakkar KN, Polli FE, Joseph RM, Tuch DS, Hadjikhani N, Barton JJS, Manoach DS. 2008 Response monitoring, repetitive behaviour and anterior cingulate abnormalities in autism spectrum disorders (ASD). *Brain* **131**, 2464–2478. (doi:10.1093/brain/awn099)

58. Ide JS, Shenoy P, Yu AJ, Li C-SR. 2013 Bayesian prediction and evaluation in the anterior cingulate cortex. *J. Neurosci.* **33**, 2039–2047. (doi:10.1523/JNEUROSCI.2201-12.2013)

59. Einarsson EÖ, Nader K. 2012 Involvement of the anterior cingulate cortex in formation, consolidation, and reconsolidation of recent and remote contextual fear memory. *Learn. Mem.* **19**, 449–452. (doi:10.1101/lm.027227.112)

60. van Velzen LS, Schmaal L, Milaneschi Y, van Tol M-J, van der Wee NJA, Veltman DJ, Penninx BWJH. 2017 Immunometabolic dysregulation is associated with reduced cortical thickness of the anterior cingulate cortex. *Brain Behav. Immun.* **60**, 361–368. (doi:10.1016/j.bbi.2016.10.019)

61. Halder R *et al.* 2016 DNA methylation changes in plasticity genes accompany the formation and maintenance of memory. *Nat. Neurosci.* **19**, 102–110. (doi:10.1038/nn.4194)

62. Sharma R *et al.* 2022 Maternal-fetal stress and DNA methylation signatures in neonatal

saliva: An epigenome-wide association study. *Research Square*. (doi:10.21203/rs.3.rs-1348030/v1)

63. Huang Z, Wang Y, Hu G, Zhou J, Mei L, Xiong W-C. 2016 YAP Is a Critical Inducer of SOCS3, Preventing Reactive Astrogliosis. *Cereb. Cortex* **26**, 2299–2310. (doi:10.1093/cercor/bhv292)

64. Abd El Gayed EM, Rizk MS, Ramadan AN, Bayomy NR. 2021 MRNA expression of the CUB and Sushi multiple domains 1 (CSMD1) and its serum protein level as predictors for psychosis in the familial high-risk children and young adults. *ACS Omega* **6**, 24128–24138. (doi:10.1021/acsomega.1c03637)

65. Liu Y, Fu X, Tang Z, Li C, Xu Y, Zhang F, Zhou D, Zhu C. 2019 Altered expression of the CSMD1 gene in the peripheral blood of schizophrenia patients. *BMC Psychiatry* **19**, 113. (doi:10.1186/s12888-019-2089-4)

66. Cukier HN *et al.* 2014 Exome sequencing of extended families with autism reveals genes shared across neurodevelopmental and neuropsychiatric disorders. *Mol. Autism* **5**, 1. (doi:10.1186/2040-2392-5-1)

67. Guo H *et al.* 2017 Genome-wide copy number variation analysis in a Chinese autism spectrum disorder cohort. *Sci. Rep.* **7**, 44155. (doi:10.1038/srep44155)

68. Deciphering Developmental Disorders Study. 2015 Large-scale discovery of novel genetic causes of developmental disorders. *Nature* **519**, 223–228. (doi:10.1038/nature14135)

69. Hoelper D, Huang H, Jain AY, Patel DJ, Lewis PW. 2017 Structural and mechanistic insights into ATRX-dependent and -independent functions of the histone chaperone DAXX. *Nat. Commun.* **8**, 1193. (doi:10.1038/s41467-017-01206-y)

70. Rubin AN, Malik R, Cho KKA, Lim KJ, Lindtner S, Robinson Schwartz SE, Vogt D, Sohal VS, Rubenstein JLR. 2020 Regulatory Elements Inserted into AAVs Confer Preferential Activity in Cortical Interneurons. *eNeuro* **7**. (doi:10.1523/ENEURO.0211-20.2020)

71. Han VX, Patel S, Jones HF, Dale RC. 2021 Maternal immune activation and neuroinflammation in human neurodevelopmental disorders. *Nat. Rev. Neurol.* **17**, 564–579. (doi:10.1038/s41582-021-00530-8)

72. Jin Y, Kong J. 2016 Transcutaneous Vagus Nerve Stimulation: A Promising Method for Treatment of Autism Spectrum Disorders. *Front. Neurosci.* **10**, 609. (doi:10.3389/fnins.2016.00609)

73. Lobmaier SM *et al.* 2020 Fetal heart rate variability responsiveness to maternal stress, non-invasively detected from maternal transabdominal ECG. *Arch. Gynecol. Obstet.* **301**, 405–414. (doi:10.1007/s00404-019-05390-8)

74. Antonelli MC, Frasch MG, Rumi M, Sharma R, Zimmermann P, Molinet MS, Lobmaier SM.

2022 Early Biomarkers and Intervention Programs for the Infant Exposed to Prenatal Stress. *Curr. Neuropharmacol.* **20**, 94–106. (doi:10.2174/1570159X19666210125150955)

75. Cheyuo C, Jacob A, Wu R, Zhou M, Coppa GF, Wang P. 2011 The parasympathetic nervous system in the quest for stroke therapeutics. *J. Cereb. Blood Flow Metab.* **31**, 1187–1195. (doi:10.1038/jcbfm.2011.24)

76. Cao J, Lu K-H, Powley TL, Liu Z. 2017 Vagal nerve stimulation triggers widespread responses and alters large-scale functional connectivity in the rat brain. *PLoS One* **12**, e0189518. (doi:10.1371/journal.pone.0189518)

77. Cao B, Wang J, Shahed M, Jelfs B, Chan RHM, Li Y. 2016 Vagus Nerve Stimulation Alters Phase Synchrony of the Anterior Cingulate Cortex and Facilitates Decision Making in Rats. *Sci. Rep.* **6**, 35135. (doi:10.1038/srep35135)

78. Andersson U, Tracey KJ. 2012 Reflex principles of immunological homeostasis. *Annu. Rev. Immunol.* **30**, 313–335. (doi:10.1146/annurev-immunol-020711-075015)

79. Frasch M, Prout A, Szynkaruk M, Gagnon R, Richardson B. 2009 Cholinergic anti-inflammatory pathway mechanisms may be active in the pre-term ovine fetus. *Reprod. Sci.* **16**, 137A.

80. Frasch MG, Nygard K, Vittal P, Zhao L, Regnault TR, Richardson BS. 2010 Translocation of neuronal high-mobility group box1 protein in relation to microglial activation in fetal sheep following repetitive umbilical cord occlusions with severe hypoxic-acidemia. *Reprod. Sci.* **17**, Accepted for Society for Gynecological Investigation meeting.

81. Nygard K, Vittal P, Richardson BS, Frasch MG. 2011 Fetal cholinergic anti-inflammatory pathway and the neuronal and astrocytic high-mobility group box 1 (HMGB1) protein release during cerebral inflammatory response. *Reprod. Sci.* **17(3) (Suppl)**, 51A.

82. Furukawa S, Sameshima H, Yang L, Ikenoue T. 2013 Activation of acetylcholine receptors and microglia in hypoxic-ischemic brain damage in newborn rats. *Brain Dev.* **35**, 607–613. (doi:10.1016/j.braindev.2012.10.006)

83. Furukawa S, Sameshima H, Yang L, Ikenoue T. 2011 Acetylcholine receptor agonist reduces brain damage induced by hypoxia-ischemia in newborn rats. *Reprod. Sci.* **18**, 172–179. (doi:10.1177/1933719110385129)

84. Su J, Yoon B-J, Dougherty ER. 2009 Accurate and reliable cancer classification based on probabilistic inference of pathway activity. *PLoS One* **4**, e8161. (doi:10.1371/journal.pone.0008161)

85. Wang Y, Jeong H, Yoon B-J, Qian X. 2020 ClusterM: a scalable algorithm for computational prediction of conserved protein complexes across multiple protein interaction networks. *BMC Genomics* **21**, 615. (doi:10.1186/s12864-020-07010-1)

86. Maddouri O, Qian X, Yoon B-J. 2021 Deep graph representations embed network information for robust disease marker identification. *Bioinformatics* (doi:10.1093/bioinformatics/btab772)